\documentclass[12pt,letterpaper]{article}

\parskip=\medskipamount            
\def\7#1#2{\mathop{\null#2}\limits^{#1}}        
\def\ast{\displaystyle *}

\def\beee{\begin{equation}}
\def\eeee{\end{equation}}
\def\dggg{^{\dagger}}

\begin{document}

\bibliographystyle{unsrt}
\begin{center}
\textbf{WHY IS CPT FUNDAMENTAL?}\\
[5mm]
O.W. Greenberg\footnote{email address, owgreen@physics.umd.edu.}\\
{\it Center for Theoretical Physics\\
Department of Physics \\
University of Maryland\\
College Park, MD~~20742-4111}\\
University of Maryland Preprint PP-04004\\
~\\

\end{center}

\begin{abstract}

G. L\"uders and W. Pauli proved the $\mathcal{CPT}$ theorem based on
Lagrangian quantum field theory almost half a century ago.
R. Jost gave a more general proof
based on ``axiomatic'' field theory nearly as long ago.
The axiomatic point of view has two advantages over the Lagrangian one.
First, the axiomatic point of view makes clear why $\mathcal{CPT}$ is
fundamental--because it is intimately related to Lorentz invariance.
Secondly, 
the axiomatic proof gives a simple way to calculate the $\mathcal{CPT}$
transform of any relativistic field without calculating $\mathcal{C}$,
$\mathcal{P}$ and $\mathcal{T}$ separately and then multiplying them.
The purpose of this pedagogical paper is to ``deaxiomatize'' the 
$\mathcal{CPT}$ theorem by explaining it in a few simple steps. We use
theorems of distribution theory and of several complex variables
without proof to make the exposition elementary.

\end{abstract}

\section{Introduction}

The notion of $\mathcal{CPT}$ symmetry, where
$\mathcal{C}$ is charge conjugation, $\mathcal{P}$ is parity (space
inversion) and $\mathcal{T}$ is time reversal in the sense of
Wigner,\footnote{$\mathcal{C}$ and $\mathcal{P}$ are unitary; 
$\mathcal{T}$ and $\mathcal{CPT}$ are antiunitary} as a symmetry
that holds for any relativistic quantum field theory evolved
from the observation of G. L\"uders~\cite{lud} that charge
conjugation symmetry and spacetime inversion symmetry both impose
the same constraints on the form of the interaction Hamiltonian
so that $\mathcal{CPT}$ symmetry has a more fundamental basis than
either $\mathcal{C}$, $\mathcal{P}$ or $\mathcal{T}$. W. Pauli~\cite{pau} 
gave a clear formulation of $\mathcal{CPT}$ symmetry in the context of 
conditions on the interaction Hamiltonian or Lagrangian. Pauli's formulation
is the form of the $\mathcal{CPT}$ symmetry that is usually discussed, the
``Lagrangian $\mathcal{CPT}$ theorem.'' R. Jost~\cite{jos}
gave a general proof of $\mathcal{CPT}$ symmetry based on the fact
that spacetime inversion is connected to the identity in the
complex Lorentz group although this inversion is not connected
to the identity in the real Lorentz group. Jost's analysis is usally
called the ``axiomatic $\mathcal{CPT}$ theorem.'' J. Schwinger~\cite{sch} 
discussed the $\mathcal{CPT}$ and spin-statistics theorems from the point
of view of his differential action principle.

Jost's proof has
been labeled as belonging to axiomatic field theory as though
that made his proof both incomprehensable and of no practical
value. In fact Jost's proof is easy to understand if one is
willing to accept theorems about distributions and analytic functions
of several complex variables without proof. 
To make notation simple in this pedagogical
paper I have exorcised all the test functions that usually appear
in discussions of singular functions (distributions). To make clear
my lack of rigor I have used the word ``analytic'' rather than
the word ``holomorphic'' in connection with the functions of several
complex variables that appear.

Jost's proof has the practical value 
that it gives a very simple and general result for the $\mathcal{CPT}$
symmetry acting on any relativistic quantum field. Jost discussed 
the $\mathcal{CPT}$ theorem in
three publications,
his original paper in Helvetica Physica Acta~\cite{jos}, his contribution
(in German) to 
the Pauli memorial volume~\cite{jos2} and his book~\cite{jos3}
on quantum field theory. Jost's theorem also is discussed in the books
by R. Streater and A.S. Wightman~\cite{str}, N.N. Bogoliubov, A.A.
Logunov and I.T. Todorov~\cite{bog}, and R. Haag~\cite{haa}.

The standard textbooks of quantum field theory all get to the
$\mathcal{CPT}$ theorem by calculating each symmetry and then calculating
their product. This is not incorrect (except for the technical fact
that each of $\mathcal{C}$, $\mathcal{P}$ and $\mathcal{T}$ can have an arbitrary
phase since they are not connected to the identity while $\mathcal{CPT}$,
which is connected to the identity, cannot have an arbitrary phase).
However calculating $\mathcal{CPT}$ by multiplying each of the three
discrete symmetries is a very complicated way to calculate $\mathcal{CPT}$.
More important, calculating $\mathcal{CPT}$ in that way
obscures why $\mathcal{CPT}$ is fundamental but none
of the individual symmetries is.

The purpose of this expository note is to explain why $\mathcal{CPT}$
is fundamental and to calculate it for a general relativistic
quantum field without worrying about the mathematical issues connected
with functions of several complex variables and their relation to
tempered distributions whose support in momentum space lies in or
on a cone. $\mathcal{CPT}$ is fundamental because it is closely related
to Lorentz covariance. We will pay attention to how far we can get with 
Lorentz covariance alone and where we must use another property of the theory.
The reader will also see that the calculation of $\mathcal{CPT}$
using general arguments is greatly simpler than the pedestrian
calculation of $\mathcal{C}$,
$\mathcal{P}$ and $\mathcal{T}$ separately and then multiplying them.
To make this note self contained we will explain ideas 
connected with group theory and field theory that many readers will
already understand. Those readers are encouraged to skip the 
introductory explanations and go directly to the $\mathcal{CPT}$
theorem itself.

\section{Representations of the real and complex Lorentz groups}

Since the heart of the argument is the fact that the connected
component of the complex Lorentz group, $L(C)$, which is the proper complex
Lorentz group, $L_+(C)$, contains spacetime inversion, we will discuss the
Lorentz group first~\cite{jos3}. 
The (real) Lorentz group can be taken as the group, $SO(1,3)$,
of real 4 x 4 matrices $\Lambda$ that preserve the metric $g$
that we take to have the form $g=diag (1, -1, -1, -1)$, 
\beee
\Lambda^T g \Lambda = g.            \label{lg}
\eeee
You can check that this condition is equivalent to saying that
a Lorentz transformation preserves the scalar product
$x^2=x \cdot x = (x^0)^2 -\sum_1^3 (x^i)^2$, i.e., 
$\Lambda x \cdot \Lambda =x \cdot x$. By taking the determinant of
Eq.(\ref{lg}) we see that $det \Lambda = \pm 1$. By looking at the
$00$ element of Eq.(\ref{lg}) we find 
$(\Lambda^0~_0)^2-\sum_1^3(\Lambda^0~_i)^2=1$, so either 
$\Lambda^0~_0 \geq 1$ or $\Lambda^0~_0 \leq -1$.
Thus the Lorentz group
falls into four disconnected components, $L^{\uparrow}_+$, $L^{\downarrow}_+$,
$L^{\uparrow}_-$, and $L^{\downarrow}_-$
 according to the sign of
the determinent of $\Lambda$ and the sign of $\Lambda^0~_0$. Only the first
of these is a group since only $L^{\uparrow}_+$ contains the identity.
We use $x \in V_+$ if $x^2 > 0, x^0 >0$; $x \in V_-$ if $x^2 > 0, x^0 <0$,
$x \sim 0$ if $x^2<0$. We also have to
consider the complex Lorentz group, the group of complex 4 x 4
matrices that obey Eq.(\ref{lg}). For the complex Lorentz group the sign
of the determinent still cannot be changed continuously, but the
matrix $-1$ is now connected to the identity, so there are only two
disconnected components. The easiest way to find the continuous family of
complex Lorentz transformations that connect the matrices $1$ and $-1$ 
is by considering the covering groups of the real and complex Lorentz 
groups, to which we now turn.

We are familiar with the fact that a spin 1/2 state transforms 
under a rotation by an angle $\theta$
with a phase $\theta/2$ rather than the phase $\theta$ of a scalar
state. So a rotation by $2\pi$ changes the phase of a spin 1/2 state
even though such a rotation should be equivalent to the identity.
Thus a spin 1/2 state does not transform as a true 
representation of the rotation group, but rather as a representation
up to a factor. The idea of a covering group is to find a larger group whose
representations are true representations without additional phases.
For the rotation group the covering group is $SU(2)$, the group of
2 x 2 unitary complex matrices with determinant 1. For the connected
component of the Lorentz
group the covering group is $\bar{L}^{\uparrow}_+ \equiv SL(2,C)$, 
the group of 2 x 2 complex
matrices of determinant 1. 

We introduce the two fundamental 
representations of $SL(2,C)$ as 
\beee
u^{\prime}_{\alpha}=A_{\alpha \beta}u_{\beta}
\eeee
and 
\beee
\dot{v}^{\prime}_{\dot{\alpha}}=
A^{\ast}_{\dot{\alpha}\dot{\beta}}\dot{v}_{\dot{\beta}},
\eeee
where $A \in SL(2,C)$ and $\ast$ stands for complex conjugate. 
Spinors with undotted
and dotted indices were introduced in~\cite{van}.
We can introduce a scalar product for these representations using the
$2 \times 2$ antisymmetric 
Levi-Civita symbol $\epsilon_{\alpha \beta}$ for the
undotted spinors and $\epsilon_{\dot{\alpha} \dot{\beta}}$ for the
dotted spinors. We choose 
$\epsilon_{12}=1,~ \epsilon_{\dot{1}\dot{2}}=1$. Any representation of
$SL(2,C)$ has the form of a spinor with $k$ undotted and $l$ dotted
indices, each transforming as given above. The only way we can reduce
these representations is by contracting with the $\epsilon$'s just
described, so the irreducible representations of $SL(2,C)$ are spinors
with $k$ symmetrized undotted and $l$ symmetrized dotted
indices. Each index corresponds to spin 1/2 so these spinors have
spin $k/2$ and $l/2$ under the $SU(2) \otimes SU(2)$ formed by taking
the groups whose generators are $J \pm iK$, where $J$ are the rotation
generators of the real Lorentz group and $K$ are the generators of
pure Lorentz transformations (boosts). (See the appendix for this
description.)

We take the Pauli matrices
to have one undotted and one dotted index, 
$(\sigma_{\mu})_{\alpha}^{\dot{\beta}}$, where $\sigma_0$ is the unit
$2 \times 2$ matrix and $\sigma_i$ are the usual Pauli matrices. Then
we can uniquely associate a $2 \times 2$ hermitian matrix $X$ 
with a real vector $x^{\mu}$ by 
$(X)_{\alpha}^{\dot{\beta}}=
x^{\mu}(\sigma_{\mu})_{\alpha}^{\dot{\beta}}$. To invert this, trace
with the $\sigma$ matrices. The reader should check that
$det X=x^2$. Recalling that matrices in $SL(2,C)$ have
determinant 1, we see that $X^{\prime} = A X A^{\dagger}$ is again
Hermitian and is 
a Lorentz transformation on $x$, where $\dagger$ stands for adjoint.
The matrices $A$ and $-A$ stand for the same Lorentz transformation;
thus the group $SL(2,C)$ covers the connected component of the Lorentz
group twice.

To cover the complex Lorentz group we allow two independent $SL(2,C)$
matrices to enter so that $X^{\prime} = A X B^T$. This $X^{\prime}$ 
is no longer Hermitian, but it still has the same Minkowski metric
length, so the covering group of the complex Lorentz group, $L_+(C)$, is 
$SL(2,C) \otimes SL(2,C)$. Since now we have
two independent matrices $A$ and $B$ at our disposal, we can achieve
$x \rightarrow -x$ either by choosing $A=1,~B=-1$ or $A=-1,~B=1$.
We can go continuously from the identity $A=1,~B=1$ to $A=1,~B=-1$
in the first case by choosing 
$A=1,~B(\phi)=diagonal(exp i \phi/2, exp -i \phi/2)$. We can find the
$4 \times 4$ complex Lorentz transformations $\Lambda(\phi)$ from
the definition of $X^{\prime}$. The result, which is a continuous
family of complex Lorentz transformations going from the identity to
spacetime inversion, is
\[
\left( \begin{array}{l}
x^{\prime 0} \\ x^{\prime 3} \\ x^{\prime 1} \\ x^{\prime 2}
\end{array} \right)
\left( \begin{array}{cccc}
cos \frac{\phi}{2}  & i~sin \frac{\phi}{2}  &      0               &  0  \\
i~sin \frac{\phi}{2} & cos \frac{\phi}{2}   & 0                    & 0 \\
0                   &    0                 &  cos \frac{\phi}{2}   & -sin \frac{\phi}{2}  \\
0                   &    0                 &   sin \frac{\phi}{2}  & cos \frac{\phi}{2}
\end{array} \right)
\left( \begin{array}{r}
x^{0} \\ x^{3} \\ x^1 \\ x^2
\end{array}  \right)
\]
Are there other ways to achieve spacetime inversion? 
In $SL(2,C)\otimes SL(2,C)$ we need 
$A X B^T = -X$, or $A X=-X B^{T~-1}$. Thus we need this relation 
where $X$ is replaced by
each of the Pauli matrices $\sigma_{\mu}$. For $\sigma_0=1$ we need 
$A=-B^{T-1}$. Then we
need $A X=X A$ where for $X$ we can choose any of the space $\sigma$'s. 
This requires $A= \omega 1$
and for $A \in SL(2,C)$ we need $\omega^2=1$ or $\omega = \pm 1$. Thus the 
\textit{only} possibilities to invert $x^{\mu}$ are the ones given above.
Now we have the group theory we need to discuss 
the $\mathcal{CPT}$ theorem.

\section{Vacuum matrix elements of products of fields define 
analytic functions.}

Next we have to discuss vacuum matrix elements of products of fields,
often called Wightman functions or distributions.  Let $\phi^{(k,l)}(x)$ 
be a field with $k$ undotted and $l$ dotted
indices, each set symmetrized, that transforms as the irreducible 
representation of $SL(2,C)$ described above. We will use the active
point of view in which a 
Poincar\'e transformation $(a,A)$ acts as\footnote{We are using the 
covering group of the Poincar\'e group, so in $(a,\Lambda)$ we replaced
$\Lambda \in L^{\uparrow}_+$ by $A \in SL(2,C)$. In the argument of
the fields on the right hand side on the next equation we replaced 
$\Lambda$ by $\Lambda(A)$ where $\Lambda(A) \in L_+^{\uparrow}$ is the 
homomorphic image
of $A \in SL(2,C)$. Where we use the the covering group of the complex Lorentz group
we should replace $\Lambda \in L_+(C)$ by $\Lambda(A,B)$ where $\Lambda(A,B)$
is the homomorphic image of $(A,B) \in SL(2,C) \otimes SL(2,C)$. To simplify
notation we will write $\Lambda$ instead of $\Lambda(A)$ or 
$\Lambda(A,B)$ in both cases.}
\beee
U(a,A)\phi^{(k,l)}(x)U(a,A)\dggg
=S^{(k,l)}(\Lambda)^{(-1)}\phi^{(k,l)}(\Lambda x+a);      \label{lc}
\eeee
The only case for which we need the detailed
form of $S^{(k,l)}(A,B)$ is when $\Lambda \in L_+(C)$ produces spacetime
inversion and for that case $S^{(k,l)}(A,B)$ is just a diagonal phase.
Thus the detailed form of $S^{(k,l)}$ is not necessary here. 
For this reason we have suppressed the indices belonging to the matrices 
$S^{(k,l)}$ as well as the indices
belonging to the field $\phi^{(k,l)}(x)$.\footnote{See the appendix for
the transformations with all indices exhibited.}
We assume the vacuum $|0\rangle$
is invariant under Poincar\'e transformations, 
\beee 
U(a,A)|0\rangle = |0\rangle.
\eeee
We were tempted first to use scalar fields in discussing Jost's proof in order
to avoid cumbersome notation and then to give the argument again for the
general case. Instead, in order to make clear how simple Jost's argument
is, we decided to streamline the notation instead and give the general case
directly. (For some properties such as the support in momentum space which
does not depend on the spin we will use the scalar case to illustrate the 
issue.) Let the single index $(p)$ (for ``pair''
of indices) stand for $(k,l)$. We will use $(p)$ and $(k,l)$ interchangeably
to label fields and other objects. 
Then the general field becomes $\phi^{(p)}(x)$, the 
matrices are $S^{(p)}(A)$, and the transformation law, again suppressing
indices, is 
\beee
U(a,A)\phi^{(p)}(x)U\dggg(a,A)
=S^{(p)}(A)^{-1}\phi^{(p)}(\Lambda x+a).      \label{ls}
\eeee
Next we write the vacuum matrix element of an arbitrary product of fields and 
use this transformation law to find
\begin{eqnarray}
\lefteqn{\langle 0| \phi^{(p_1)}(x_1) \phi^{(p_2)}(x_2) \cdots  
\phi^{(p_n)}(x_n) |0\rangle}   \nonumber \\
&  & = ( |0\rangle, \phi^{(p_1)}(x_1) \phi^{(p_2)}(x_2) 
\cdots  \phi^{(p_n)}(x_n) |0\rangle )  \nonumber \\
&  & =(U(a,A)|0\rangle, U(a,A) 
\phi^{(p_1)}(x_1) \phi^{(p_2)}(x_2) \cdots  \phi^{(p_n)}(x_n) |0\rangle )   \nonumber \\
&  & =(|0\rangle,[\prod^n_1S^{(p_i)}(A)^{-1}] 
\phi^{(p_1)}(\Lambda x_1 + a) \phi^{(p_2)}(\Lambda x_2 + a) \cdots \phi^{(p_n)}(\Lambda x_n + a)
|0\rangle )  \nonumber\\
&  & =[\prod^n_1S^{(p_i)}(A)^{-1}]\langle 0|\phi^{(p_1)}(\Lambda x_1 + a) 
\phi^{(p_2)}(\Lambda x_2 + a) \cdots \phi^{(p_n)}(\Lambda x_n + a)
|0\rangle .
\end{eqnarray}
Because of translation invariance this matrix element depends on only $n-1$ differences
of the spacetime coordinates. We define the Wightman function~\cite{wig} which is a
generalized function or distribution
\beee
W^{(n;p_1p_2 \cdots p_n)}(x_1-x_2, x_2-x_3, \cdots , x_{n-1}-x_n) \equiv
\langle 0| \phi^{(p_1)}(x_1) \phi^{(p_2)}(x_2) \cdots  \phi^{(p_n)}(x_n) |0\rangle.
\eeee
Since we will have to deal with three kinds of difference vectors, we will
use different letters to distinguish them: $\xi$ for real vectors, 
$\rho$ for real vectors, called ``Jost points'' defined below, in the
domain of analyticity $\mathcal{T}_{n-1}^{\prime}$, 
and $\zeta$ for complex vectors.
To streamline the notation we compress the indices $(p_1p_2 \cdots p_n)$ on
$F^{(n)}$ to a single index $(\wp)$.
We define $\xi_j=x_j-x_{j+1}$. Then invariance under the connected component of the
Lorentz group (the proper orthochronous component, $L_+^{\uparrow})$ gives
\beee
W^{(n;\wp)}(\Lambda \xi_1, \Lambda \xi_2, \cdots , \Lambda \xi_{n-1}) =
[\prod^n_1S^{(p_i)}(A)]W^{(n;\wp)}(\xi_1, \xi_2, \cdots ,  \xi_{n-1}).
\eeee
The requirement that physical states have positive energy, except the vacuum which has
zero energy, implies that the momenta in the Fourier transform of the $F^{(n;\wp)}$'s lie
in or on the forward light cone. To see this we drop all indices and 
consider a matrix element of a scalar field, since the support in momentum
space depends only on the translation subgroup of the Poincar\'{e} group, 
\beee
\langle 0|\phi(x_1) \cdots \phi(x_j)U(a,1)\phi(x_{j+1})\cdots \phi(x_n)|0 \rangle.
\eeee
Evaluate this either by letting the translation operator act to the right
to get
\beee
W^{(n)}(\xi_1, \cdots , \xi_{j-1}, \xi_j +a, \xi_{j+1}, \cdots \xi_{n-1}),
\eeee
or by inserting the identity operator using a complete set of intermediate 
states, $|q_j,\alpha_j\rangle$, before the translation operator to get
\beee
\langle 0|\phi(x_1) \cdots \phi(x_j) \sum_{q_j,\alpha_j} |q_j,\alpha_j \rangle
\langle q_j,\alpha_j|exp(-iq_j \cdot a) \phi(x_{j+1}) \cdots \phi(x_n)|0\rangle.
\eeee
This shows that the physical momenta $q_j$ are conjugate to the spacetime difference
vectors $\xi_j$. Thus
\beee
W^{(n)}(\xi_1, \xi_2, \cdots ,  \xi_{n-1})=\frac{1}{(2\pi)^{4(n-1)}}
\prod_1^{n-1} \int d^4q_j exp(-i q_j \cdot \xi_j) \tilde{W}^{n}(q_1, \cdots q_{n-1}),
\eeee
where the momenta $q_j$ are physical, i.e., are in or on the forward light cone.

Now use the intuitive criterion that a Fourier transform that is a distribution
becomes an analytic function when the external variable is made complex in such
a way as to provide a damping factor so that the Fourier transform becomes a 
Laplace transform. Thus we must examine when the factor $exp(- iq_j \cdot \xi_j)$
becomes decreasing for $\xi_j \rightarrow \zeta_j=\xi_j + i\eta_j$ with $\xi_j$
and $\eta_j$ real vectors. What counts is the absolute value of the factor which
is $exp(q_j \cdot \eta_j)$. This becomes decreasing if $\eta_j$ is in or on
the backward
light cone since the physical momentum $q_j$ is in the forward light cone. 
Thus the Wightman function
\beee
W^{(n)}(\zeta_1, \zeta_2, \cdots ,  \zeta_{n-1})
\eeee
is an analytic function of $4(n-1)$ complex variables (in four-dimensional spacetime)
when $Im\zeta_j \in V^{\uparrow}_-$. It is also single-valued.

\section{Enlargement of the domain of analyticity}

As it stands, $W^{(n)}$ is analytic
only when $Im \zeta_i \neq 0$, i.e., its domain of analyticity has no real points.
Call this domain, which has the form of a tube with $Re \zeta_i$ arbitrary and 
$Im \zeta_i in a  \in V_-$, the tube $\mathcal{T}_{n-1}$. 
Now we restore the labels of the fields and 
use a profound result due to V. Bargmann, D.W. Hall and 
A.S. Wightman~\cite{bar} which applies to this case. If 
\beee
W^{(n;\wp)}(\Lambda\zeta_1,\Lambda \zeta_2, \cdots ,\Lambda \zeta_{n-1})=
[\prod^n_1S^{(p_i)}(A)]
W^{(n;\wp)}(\zeta_1, \zeta_2, \cdots ,  \zeta_{n-1})    \label{cov}
\eeee
for the covering group, $SL(2,C)$, of real $\Lambda \in L^{\uparrow}_+$ then 
$W^{(n;\wp)}(\zeta_1, \zeta_2, \cdots ,  \zeta_{n-1})$ has a (unique) single-valued
analytic continuation to the domain $\mathcal{T}_{n-1}^{\prime}$, 
that we call the
extended tube,\footnote{although technically it is not a tube in the variables
$(\zeta_1, \cdots, \zeta_{n-1})$}
that is the union of
all $\Lambda \mathcal{T}_{n-1}$ where now we have complex matrices
$\Lambda \in L_+(C)$. 
This enlargement of the domain of analyticity leads to 
two crucial results. First, in contrast
to $\mathcal{T}_{n-1}$, our new, larger domain of analyticity, 
$\mathcal{T}_{n-1}^{\prime}$, contains real points of analyticity $(\rho_j)$ that
we will discuss below. Secondly since
$\mathcal{T}_{n-1}^{\prime}$ is invariant under complex Lorentz transformations in
$L_+(C)$, one of which is spacetime inversion, we have the relation
\beee
W^{(n;\wp)}(\zeta_1, \zeta_2, \cdots ,  \zeta_{n-1})=
(-1)^L W^{(n;\wp)}(-\zeta_1, -\zeta_2, \cdots ,  -\zeta_{n-1})
\eeee                 \label{inversion}
$L=\sum l_i$,
in $\mathcal{T}_{n-1}^{\prime}$. To see where the factor $(-1)^L$ comes from 
we repeat that for $\Lambda \in L^{\uparrow}_+$, $\Lambda$ depends on the 
matrices $A$ and $A^*$ in $SL(2,C);$ however, as mentioned in a footnote
above, now that we have the
extension to $L_+(C)$, $\Lambda$ depends on two independent 
matrices $A$ and $B$ in
$SL(2,C)$ and we can transform continuously from the identity
to spacetime inversion. The $S^{(k,l)}$ matrices for spacetime inversion are 
just powers of
$(-1)$; thus if we choose $A=1, ~B=-1$ then 
$S^{(k,l)}(1,-1)=(-1)^l \mathbf{1}$ \footnote{Here $\mathbf{1}$ is the 
direct product of a $(2k+1) \times (2k+1)$ unit matrix and a 
$(2l+1) \times (2l+1)$ unit matrix and 
we find the result just stated for $L$.} Note that spacetime inversion here is
an element of the complex Lorentz group and is unitary, not antiunitary.

Jost found a precise characterization of the real points, $(\rho_j)$,
in $\mathcal{T}_{n-1}^{\prime}$: 
$\sum_1^{n-1} \lambda_i \rho_i \sim 0$,
for all $\lambda_i >0$ such that $\sum_1^{n-1}\lambda_i \geq 0$. 
This requires that each $\rho_i \sim 0$. 

Jost's result is particularly simple for the $W^{(2)}$ function for 
single scalar field in which there is one
complex difference vector $\zeta$. 
Since $W^{(2)}(\Lambda \zeta)=W^{(2)}(\zeta)$ we can
find the extended tube $\mathcal{T}_{1}^{\prime}$ by finding the 
values of $\zeta^2$ that
can be obtained from $\Lambda \zeta$ with 
$\zeta = \xi + i \eta$, $\eta \in V_-$. Then
$\zeta^2=\xi^2 -\eta^2 +2i 
\xi \cdot \eta$. The real points are those for which 
$\xi \cdot \eta =0$, with $\eta \in V_-$.
These points are the spacelike points $\xi \sim 0$, in agreement with 
Jost's general result.

\section{The general formula for $\mathcal{CPT}$.}

When we write the relation between Wightman functions at 
Jost points that comes from
spacetime inversion, Eq.(\ref{inversion}), 
in terms of vacuum matrix elements we find
\begin{eqnarray} \lefteqn{
\langle 0| \phi^{(k_1,l_1)}(x_1) \phi^{(k_2,l_2)}(x_2) \cdots \phi^{(k_n,l_n)}(x_n) |0\rangle
=}  \nonumber  \\
& &
(-1)^L \langle 0| \phi^{(k_1,l_1)}(-x_1) \phi^{(k_2,l_2)}(-x_2) 
\cdots \phi^{(k_n,l_n)}(-x_n) |0\rangle.
\end{eqnarray}
Each side of this last 
equation can be analytically continued; the left hand side to complex 
$\zeta_i$ with $\eta = Im  \zeta \in V_+^{\downarrow}$ and the right hand
side to complex $\zeta_i$ with $\eta = Im  \zeta \in V_+^{\uparrow}$.
We cannot take the limit as $\eta_i \rightarrow 0$ to get a relation between 
vacuum matrix elements of products of the fields, because, as just noted, 
if $\zeta_i$ has its imaginary part
in the backward cone, then $-\zeta_i$ has its imaginary part in the forward 
cone and then the analytic continuations of the functions on the two sides of 
Eq.(\ref{inversion}) 
are not valid in the same domain. On the other hand if we 
consider the vacuum matrix element with the fields in completely reversed 
order,
\beee
\langle 0| \phi^{(p_n)}(-x_n)  \cdots  \phi^{(p_2)}(-x_2) \phi^{(p_1)}(-x_1) |0\rangle, 
\eeee
which corresponds to 
\beee
W^{(n;i\wp)}(\xi_{n-1},  \cdots , \xi_2, \xi_1),
\eeee
in terms of difference variables, where $i\wp$ stands for 
$(p_n \cdots p_2p_1)$, 
both functions will have the same domain, 
$\mathcal{T}_{n-1}^{\prime}$,
of analyticity. This is precisely where we have to assume something beyond Lorentz
covariance. To reverse the order of all the fields 
we assume that at a Jost point the two vacuum matrix elements are related by 
a sign $(-1)^I$ where $I$ is the number of transpositions of Fermi fields 
necessary to invert the order of the fields. This is implied by the 
spin-locality theorem\footnote{This condition is usually called the 
spin-statistics theorem. We have argued that in the present context it should
be called the spin-locality theorem.~\cite{sl}} but is weaker 
than that theorem since we need this relation only at Jost points
(or even only in a neighborhood of a Jost point) in each
matrix element, rather than as an operator relation. If there 
are $F$ Fermi fields then $I=(F-1)+(F-2)+ \cdots + 1=1/2 F(F-1)$. 
Since the number of Fermi fields in a nonvanishing vacuum matrix element must 
be even, $F-1$ must be even, thus the phase that enters is
$(-1)^{1/2 F(F-1)}=((-1)^{(F-1)})^{F/2}=(-1)^{F/2}=i^F$. The condition on
matrix elements that
\begin{eqnarray} \lefteqn{
\langle 0| \phi^{(k_1,l_1)}(x_1) \phi^{(k_2,l_2)}(x_2) \cdots \phi^{(k_n,l_n)}(x_n) |0\rangle
=}  \nonumber  \\
& &
i^F \langle 0| \phi^{(k_1,l_1)}(x_n) \cdots \phi^{(k_2,l_2)}(x_2) 
\phi^{(k_n,l_n)}(x_1) |0\rangle
\end{eqnarray}
at Jost points is called ``weak local commutativity.'' 
Clearly local commutativity (sometimes called microcausality) implies
weak local commutativity. 
Combining spacetime inversion and weak local
commutativity and collecting phases, we have 
\beee
W^{(n;\wp)}(\zeta_1, \zeta_2, \cdots ,  \zeta_{n-1})=
i^F (-1)^L W^{(n;i\wp)}(\zeta_{n-1},\cdots , \zeta_2, \zeta_1)
\eeee
Now we can take $Im \zeta_i \rightarrow 0$ 
on both sides and we get an equality between distributions for all 
$\xi_i$,
\beee
W^{(n;\wp))}(\xi_1, \xi_2, \cdots ,  \xi_{n-1})=
i^F (-1)^L W^{(n;i\wp)}(\xi_{n-1},  \cdots , \xi_2, \xi_1).
\eeee   
Translated back into vacuum matrix elements this says
\begin{eqnarray} \lefteqn{
\langle 0| \phi^{(p_1)}(x_1) \phi^{(p_2)}(x_2) 
\cdots  \phi^{(p_n)}(x_n) |0\rangle=}      \nonumber  \\
& & i^F (-1)^L
\langle 0| \phi^{(p_n)}(-x_n)  \cdots  \phi^{(p_2)}(-x_2) 
\phi^{(p_1)}(-x_1) |0\rangle.
\end{eqnarray}
Replacing $(p_j)$ by $(k_j,l_j)$ we have         
\begin{eqnarray} \lefteqn{
\langle 0| \phi^{(k_1,l_1)}(x_1) \phi^{(k_2,l_2)}(x_2) \cdots \phi^{(k_n,l_n)}(x_n) |0\rangle
=}  \nonumber  \\
& &
i^F(-1)^L \langle 0| \phi^{(k_n,l_n)}(-x_n) \cdots \phi^{(k_2,l_2)}(-x_2) 
\phi^{(k_1,l_1)}(-x_1) |0\rangle.
\end{eqnarray}
 
We can restore the original order of the fields on the right hand side by using the 
hermiticity of the scalar product, $(\Psi,\Xi)=(\Xi,\Psi)^{\ast}.$ The
appearance of complex conjugation is fine, 
since we know that $\mathcal{CPT}$ is antiunitary. We find
\begin{eqnarray} \lefteqn{
\langle 0| \phi^{(k_1,l_1)}(x_1) \phi^{(k_2,l_2)}(x_2) \cdots \phi^{(k_n,l_n)}(x_n) |0\rangle
=}  \nonumber  \\
& &
i^F(-1)^L \langle 0| \phi^{(k_1,l_1)\dagger}(-x_1) \phi^{(k_2,l_2)\dagger}(-x_2) \cdots
\phi^{(k_n,l_n)\dagger}(-x_n) |0\rangle^{\ast},   \label{vaccpt}
\end{eqnarray}
where $F=\sum_1^n f_i$ and $f$ is zero for a Bose field (with $k+l$ even) and 
one for a Fermi field (with $k+l$ odd).
Now we can read off what $\mathcal{CPT}$, which for brevity we call $\Theta$, must be,
\beee
\Theta  \phi^{(k,l)}(x)\Theta\dggg=(-1)^l (\pm i)^f \phi^{(k,l)\dagger}(-x).
\label{cpt}
\eeee
When we embed this relation in an arbitrary vacuum matrix element and use the invariance
of the vacuum, $\Theta |0\rangle=|0\rangle$,
we find precisely Eq.(\ref{vaccpt})! When we run this sequence of relations the other way,
we conclude that weak local commutativity in the neighborhood of a Jost point
is necessary and sufficient for $\mathcal{CPT}$. 
There is a sign ambiguity because $F$, the total number of Fermi fields in a nonvanishing
vacuum matrix element, must be even.

Note that $\mathcal{CPT}$ takes each irreducible of $L^{\uparrow}_+$ to a phase times
itself, so that for example the part $\phi^{(1,0)}$ of the Dirac spin field is mapped to
$i\phi^{(1,0)}$ and the part $\phi^{(0,1)}$ is mapped to $-i\phi^{(0,1)}$.
Both the vector and axial vector fields have the form $\phi^{(1,1)}$ so 
these fields are indistinguishable under $\Theta$ and both get the phase $(-1)$ 
under $\Theta$. The analogous statements hold for the scalar and pseudoscalar
fields, $\phi^{(0,0)}$, which both get phase $1$. The antisuymmetric rank two
tensor field $T^{\mu \nu}$
is $\phi^{(2,0)} + \phi^{(0,2)}$ and also gets phase $1$. The traceless
symmetric tensor of rank two is $\phi^{(2,2)}$ and also gets phase $1$.

The $\mathcal{CPT}$ operator $\Theta$ interchanges 
undotted and dotted indices, so that
\beee
\phi^{(k,l)\dagger}(x) = \phi^{\dagger (l,k)}(x).
\eeee
Under $\Theta$ particles and antiparticles are interchanged (some particles
may be identical to their antiparticles). Energies and momenta stay the same;
spin components and helicities are reversed. 

When we act twice by $\Theta$ we find
\begin{eqnarray}
\Theta^2 \phi^{(k,l)}(x)\Theta^{\dagger 2}  
& = & 
\Theta (-1)^l (\pm i)^f \phi^{(k,l)~\dagger}(-x)\Theta^{\dagger}  
\nonumber \\
& = & \Theta (-1)^l (\mp i)^f \Theta \phi^{(k,l)~\dagger}(-x) 
\Theta^{\dagger} \nonumber \\
& = & (-1)^f \phi^{(k,l)}(x),
\end{eqnarray}
so $\Theta^2$ commutes with Bose fields and anticommutes with Fermi fields. The
reader can check that the phase of $\Theta^2$ cannot be changed by changing
a phase in the definition of $\Theta$. This is true for all antiunitary 
operators.

\section{$\mathcal{CPT}$ for the $S$-matrix.}

Because $\Theta$ reverses time,
in and out states are interchanged. Taking the antiunitarity of $\Theta$ into
account, the $S$-matrix obeys
\beee
S_{\alpha, \beta} \equiv _{out}\langle \alpha | \beta \rangle_{in}
=_{out}\langle \hat{\beta} | \hat{\alpha} \rangle_{in}=
S_{\hat{\beta},\hat{\alpha}},
\eeee
where $|\hat{\alpha} \rangle$ has particles and antiparticles exchanged, spin 
components and helicities reversed, and energies and momenta the same as
in $|\alpha \rangle$

In terms of the $S$-operator this is
\beee
\Theta S \Theta\dggg = S^{-1}, \mathrm{or}~~\Theta S  = S^{-1} \Theta.
\eeee

\section{Summary.}

Jost's general proof of the $\mathcal{CPT}$ theorem leads directly to
a definition of the $\mathcal{CPT}$ symmetry applied to fields
belonging to an arbitrary irreducible representation of $SL(2,C)$, the 
covering group of the real proper orthochronous Lorentz group, 
$L^{\uparrow}_+$. In the real Lorentz group spacetime inversion is not
connected to the identity; however in the complex Lorentz group 
$x \rightarrow -x$ is connected to the identity. Using the Wightman
analytic functions that are analytic continuations of vacuum matrix elements
of products of the fields, the Bargmann-Hall-Wightman theorem allows
analytic continuation of the Wightman functions from (the covering group)
of) $L^{\uparrow}_+$ to (the covering group of) the complex Lorentz group
$L_+(C)$ which now allows spacetime inversion. 
In the larger domain of analyticity given by the 
Bargmann-Hall-Wightman theorem there are real points of analyticity, the
Jost points; however we cannot take limits to get a general 
relation between the
the matrix elements at the original points and at the spacetime inverted
points. However, if we invert the order of the fields, we can get the general 
relation. This is the only step where we must assume something beyond 
Lorentz covariance: we must
assume weak local commutativity at Jost points to allow the reordering.
We can restore the original order of the fields by using the hermiticity of
the scalar product, which is not an additional assumption. This step complex
conjugates the matrix elements which means, as we expect, that $\mathcal{CPT}$
is antiunitary rather than unitary. Now we are able to read off the general,
simple result~Eq. (\ref{cpt}) for $\mathcal{CPT}$ on each irreducible of 
(the covering group of) the Lorentz group, $L^{\uparrow}_+$.

\appendix

\section{Alternative description of the irreducible representations of the
Lorentz group}

An alternative way to describe the irreducible representations of the
Lorentz group is to combine the rotation generators, $J_i$, with the
boost (pure Lorentz transformation) generators, $K_i$, into a pair of
commuting $SU(2)$ generators $A_i$ and $B_i$ . The irreducible fields
are $\psi^{(A,B)}$ where $A$ and $B$ are the spins associated with each
of the $SU(2)$ algebras~\cite{wei}. The relation between these two description is
$\phi^{(2A,2B)}=\psi^{(A,B)}$
The reader can check that the adjoint of $\phi^{(k,l)}$ transforms
as a field $\chi^{(l,k)}$; that is the dotted and undotted indices get
interchanged. For this reason we don't have to talk about adjoints of the 
fields; they are taken care of if we allow an arbitrary irreducible--they don't 
introduce anything new.

\section{Detailed form of the transformation law for the fields}
\begin{eqnarray} \lefteqn{
 U(a,A)\phi^{(k,l)}_{\alpha_1 \cdots \alpha_k \dot{\beta}_1 \cdots
 \dot{\beta}_l}(x)U\dggg(a,A)
 =}  \nonumber  \\
 & & 
 A^{-1}_{\alpha_1\alpha^{\prime}_1} \cdots 
 A^{-1}_{\alpha_k\alpha^{\prime}_k}
 A^{-1\ast}_{\dot{\beta}_1\dot{\beta}^{\prime}_1} \cdots 
 A^{-1\ast}_{\dot{\beta}_l\dot{\beta}^{\prime}_l}
 \phi^{(k,l)}_{\alpha^{\prime}_1 \cdots \alpha^{\prime}_k
 \dot{\beta}^{\prime}_1 \cdots
 \dot{\beta}^{\prime}_l}(\Lambda(A) x+a),     
 \end{eqnarray}
 for $\Lambda(A) \in L^{\uparrow}_+$, and 

\newpage

 \begin{eqnarray} \lefteqn{
 U(a,A,B)\phi^{(k,l)}_{\alpha_1 \cdots \alpha_k \dot{\beta}_1 \cdots
 \dot{\beta}_l}(x)U\dggg(a,A,B)
 =}  \nonumber  \\
 & &
 A^{-1}_{\alpha_1\alpha^{\prime}_1} \cdots 
 A^{-1}_{\alpha_k\alpha^{\prime}_k}
 B^T_{\dot{\beta}_1\dot{\beta}^{\prime}_1} \cdots 
 B^T_{\dot{\beta}_l\dot{\beta}^{\prime}_l}
 \phi^{(k,l)}_{\alpha^{\prime}_1 \cdots \alpha^{\prime}_k
 \dot{\beta}^{\prime}_1 \cdots
 \dot{\beta}^{\prime}_l}(\Lambda(A,B) x+a),     
 \end{eqnarray}
for $\Lambda(A,B) \in L_+(C)$,
where $\phi$ is symmetric in the $\alpha$'s and in the $\dot{\beta}$'s 
separately for both cases.

\section{Qualitative difference nature of domains of analyticity for
functions of several complex variables}

Readers
can ignore this appendix which is not necessary for our discussion of the $\mathcal{CPT}$
theorem. Functions of several complex variables differ qualitatively
from functions of a single complex variable in their possible domains
of analyticity. For a single complex variable, for every domain in the complex plane bounded by
a smooth curve there is an analytic function that cannot be continued outside this domain.
Thus any such region is the domain of analyticity for some function. For several complex
variables this is not true. Domains of analyticity must be ``holomorphically convex.'' 
Intuitively such domains must not have ``dimples'' that can be removed by analytically
continuing \textit{any} analytic function across the dimple using a Cauchy contour that
surronds the dimple. The additional dimensions available for several complex variables
is what allows this analytic continuation. This possibility of analytic continuation comes 
into play when the commutativity or anticommutativity of fields at spacelike separation is
imposed on the Wightman functions. In the case of interest for the $\mathcal{CPT}$ theorem,
which involves reversing all the fields in the vacuum matrix element, the new and old extended
tubes agree, so further analytic continuation is not possible.

\section{Lorentz covariance alone does not suffice}

Lorentz covariance alone is not sufficient for $\mathcal{CPT}$. One example
suffices to show this. A free or generalized free field can be Lorentz 
covariant but not obey $\mathcal{CPT}$ invariance if the particle and 
antiparticle masses are different~\cite{owg}. What fails
in that case is that WLC does not hold at Jost points. This possibility
is associated with the purely timelike support in momentum space of free 
or generalized free fields; for timelike momenta positive and negative energies
can be separated in a covariant way. By contrast positive and negative energies
can be transformed into each other for spacelike momenta. Note that although
the fields in these examples transform covariantly their time-ordered products
are not covariant. Thus if we require that time-ordered products be covariant
as part of Lorentz covariance of a theory then, as shown in~\cite{owg}, 
free fields that violate $\mathcal{CPT}$ are not covariant. See~\cite{owg2}
for a detailed analysis of hybrid Dirac fields 
(``homeotic'' fields~\cite{baren}) which
can be covariant only when they are non-interacting but even in the free
case have time-ordered products that are not covariant

Acknowledgement: This work
was supported in part by the National Science Foundation, 
Award No. PHY-0140301.


\begin{thebibliography}{20}

\bibitem{lud} G. L\"uders, Det. Kong. Danske Videnskabernes Selskab,
Mat.-fys. Medd., 28, No. 5 (1954).

\bibitem{pau} W. Pauli, in \textit{Niels Bohr and the Development of Physics},
(McGraw-Hill, New York, 1955), pp30-51.

\bibitem{jos} R. Jost, Helv. Phys. Acta 30, 409 (1957).

\bibitem{sch} J. Schwinger, Proc. Nat. Acad. Sci. 44, 223 (1958) and cited
references.

\bibitem{jos2} R. Jost, in \textit{Theoretical Physics in the Twentieth
Century}, (Interscience, New York, 1960). 

\bibitem{jos3} R. Jost, \textit{The General Theory of Quantized Fields},
(American Mathematical Society, Providence, 1965).

\bibitem{str} R.F. Streater and A.S. Wightman, \textit{PCT, Spin \& Statistics,
and All That}, (Benjamin, New York, 1964).

\bibitem{bog} N.N. Bogoliubov, A.A.
Logunov and I.T. Todorov, \textit{Introduction to Axiomatic Quantum Field
Theory}, (Benjamin, Reading, 1975).

\bibitem{haa} R. Haag, \textit{Local Quantum Physics}, (Springer, Berlin, 1996).

\bibitem{van} B.L. van der Waerden, \textit{Group Theory and Quantum Mechanics},
(Springer, Berlin, 1974).

\bibitem{wig} A.S. Wightman, Phys. Rev. 101, 860 (1956).

\bibitem{bar} D.W. Hall and A.S. Wightman, Det. Kong. Danske Videnskabernes 
Selskab, Mat.-fys. Medd. 31, No. 5 (1957).

\bibitem{wei} S. Weinberg, \textit{The Quantum Theory of Fields, Vol. I},
(Cambridge, Cambridge, 1995).

\bibitem{wei2} Weinberg, \textit{op. cit.}, pp 244-246.

\bibitem{sl} O.W. Greenberg, Phys. Lett. B 416, 144 (1998).

\bibitem{owg} O.W. Greenberg, Phys. Rev. Lett. 89, 231602-1 (2002).

\bibitem{owg2} O.W. Greenberg, Phys. Lett. B 567, 179 (2003).

\bibitem{baren} G. Barenboim and J. Lykken, Phys. Lett. B 554, 73(2003). 

\end{thebibliography}
\end{document}